\documentstyle[12pt]{article}
\input psfig.tex
\begin{document}
\psnoisy
\def\ps{post\-script}
\title{Exploring vector meson masses in nuclear collisions
\footnote {Talk delivered by B. K. Jain at National Seminar on Nuclear
Physics, July 26-29, 1999, Institute of Physics, Bhubaneswar, India}}
\author{B. K. Jain and Swapan  Das \\
Nuclear Physics Division, Bhabha Atomic Research Centre,\\
Mumbai 400 085, India \\
\\
Bijoy Kundu \\ Institute   of  Physics,  Sachivalaya  Marg, \\
Bhubaneswar 751 005, India}
\maketitle  \def\eqn{equation}
\centerline{\bf Abstract}
The formalism developed earlier by us
for the propagation of a resonance in the nuclear medium
in proton-nucleus collisions has been
modified to the
case of vector boson production in heavy-ion collisions.
The first part of the talk describes this formalism.
The formalism includes coherently the contribution
to the observed
di-lepton production from the decay of a
vector boson inside as well as
outside the nuclear medium.
The calculated invariant rho mass distributions are presented for
the $\rho $-meson production using optical potentials estimated
within the VDM and the resonance model.

In the second part of the talk we write a formalism for coherent
rho production in  proton  nucleus  collisions  and  explore  the
sensitivity  of the (p,p$^\prime \rho ^0$) reaction cross section
to medium mass modification of the rho meson.
\newpage

\section {Introduction}

The  modification  of  hadron  masses in the nuclear medium is an
issue of much interest currently \cite{hdms}. Since  in  QCD  the
hadrons  are  excitations of the vacuum, it is natural that these
excitations can get affected by the proximity of  other  hadrons.
Experimentally  the  masses  of  unstable hadrons are explored by
producing them  in  nuclear  reactions  and  then  measuring  the
invariant masses of their leptonic decay products \cite{dre}. The
medium  modification  of  the unstable hadron from these data is,
normally, inferred by adding incoherently the decay of the hadron
inside and outside  the  nuclear  medium  \cite{incoh}.  Recently
formalisms  have been developed by us \cite{jnku} and Boreskov et
al. \cite{bor} for the  propagation  of  resonances  produced  in
proton-nucleus   collisions.  These  formalisms  incorporate  the
interaction  of  the  resonance  with  the  nuclear  medium.  The
invariant  mass  spectrum of the measured decay products in these
formalisms is obtained by adding coherently the  contribution  of
the  resonance  decay  inside and outside the nuclear medium. The
formalism  of  Jain  et  al.  \cite{jnku}   also   includes   the
interaction  of  the  resonance  decay  products with the nuclear
medium if the latter are  hadrons.  In  the  first  part  of  the
present talk we give the modification of our earlier formalism to
heavy-ion  collisions,  and apply it to the propagation and decay
of rho-meson. Our aim is to see as how the medium modification of
$\rho  $-meson  shows  up  in  the   di-lepton   invariant   mass
distribution  using  a  proper quantum mechanical description for
the rho propagation  in  the  nucleus.  We  do  not  compare  our
calculated   results  with  the  existing  experimental  data  on
rho-meson production in high energy heavy-ion collisions, because
these data contain  contribution  from  several  other  di-lepton
processes than that considered in the present paper.

In  the second part of the talk we write a formalism for coherent
rho production in  proton  nucleus  collisions  and  explore  the
sensitivity  of the (p,p$^\prime \rho ^0$) reaction cross section
to medium mass modification of the rho meson.

\section {Heavy-ion collisions}
\subsection{Formalism}
Let  us suppose that two heavy nuclei, one the projectile $A$ and
another the target $B$, collide at high energies. We assume  that
one nucleon in the projectile and one in the target collide and a
resonance  $R$ is produced at the collision point. This resonance
then moves along the  beam  direction,  which  is  taken  as  the
z-axis,  and  decays  at some subsequent point. Denoting by ${\bf
r}$  the  relative  coordinate  between  the   target   and   the
projectile,   and   by   ${\bf  (r_{B},  r_{A})}$  the  intrinsic
coordinates of the target and projectile nucleons,  respectively,
the resonance coordinate is written as,
\begin{eqnarray}
{\bf r_R} = {\bf r_A} + \frac {B} {A+B} {\bf r}.
\end{eqnarray}
With  this  definition  the  ratio  of the cross sections for the
resonance production in AB and NN collisions,  for  an  inclusive
situation  where the state of the (A+B) system is not identified,
can be written as,
\begin{eqnarray}
\frac {\Delta \sigma _R^{AB}}{\Delta \sigma _R^{NN}} =[K.F.] \int d {\bf b}
\int dz \int d {\bf r_A} \rho_A ({\bf r_A}) \rho_B ({\bf r + r_A})
|G({\bf r_R;k_R,\mu})|^2,
\label{sigma}
\end{eqnarray}
where  $\rho_x$ are the nuclear densities. [K.F.] is the relevant
kinematical factor. The function $G({\bf r_R;k_R,\mu})$ physically
gives the probability amplitude for finding the decay products of
the resonance in the detector with the total momentum ${\bf k_R}$
and the invariant mass $\mu $, if the resonance is produced at  a
point   ${\bf   r_R}$  in  the  nucleus  (for  details  see  Ref.
\cite{jnku,bor}. In terms of  the  resonance  propagator  $G({\bf
r^{'}_R,r_R})$, the function $G({\bf r_R;k_R,\mu})$ is defined as
\begin{eqnarray}
G({\bf r_R;k_R,\mu}) = \int d{\bf r^{'}_R } exp(-i {\bf k_R.r^{'}_R})
G({\bf r^{'}_R,r_R}),
\end{eqnarray}
where $G({\bf r^{'}_R,r_R})$ satisfies
\begin{eqnarray}
[\nabla^2 + E^2 -m^2_R +i \Gamma_R m_R -\Pi_R] G({\bf r^{'}_R,r_R})=
\delta ({\bf r^{'}_R-r_R}).
\end{eqnarray}
Here  $\Pi_R$  is  the self energy of the resonance in the medium
and $\Gamma_R$ is its free space decay width.

In the eikonal approximation we can write,
\begin{eqnarray}
G({\bf r_R;k_R,\mu})=exp(-i {\bf k_R.r_R})\phi_R({\bf r_R;k_R,\mu}),
\end{eqnarray}
where $\phi_R$ is a slowly varying modulating function. With this, and
using the Eqs. (3,4), $\phi_R$ approximately
works out to
\begin{eqnarray}
\phi({\bf r_R;k_R,\mu}) &=& \frac {1} {2ik_R} \int dz^{'}_R exp[\frac {1}{2ik_R}
(\mu^2 -m^2_R+i\Gamma_R m_R)(z_R-z^{'}_R)] \nonumber \\
& &\times exp[\frac {-i}{v_R} \int ^{z'_R}_{z_R}
V_R (b_R,z{''}) d z^{''}_R].
\end{eqnarray}
Here we have written
\begin{eqnarray}
\Pi_R= 2 E_R V_R,
\end{eqnarray}
where  $V_R$ is the optical potential of the resonance, R, in the
nuclear medium. In general, it is complex. Its real part,  as  we
shall  see  later,  is related to the mass shift of the resonance
and the imaginary part gives  the  collision  broadening  of  the
resonance in the medium.

For   a   nucleus  with  a  sharp  surface,  function  $\phi({\bf
r_R;k_R,\mu})$ splits into a sum of two terms, one  corresponding
to  the  decay of the resonance inside the nucleus and another to
the decay outside the nucleus, i.e.
\begin{eqnarray}
\phi({\bf r_R;k_R,\mu}) = \phi _{in} ({\bf r_R}) + \phi_{out} ({\bf r_R})
\end{eqnarray}
with
\begin{eqnarray}
\phi_{in}({\bf r_R}) = \frac{1}{2ik_R} \int_{z_R}^{\sqrt (R^2-b^2)} d z'_R
\phi_R
({\bf b_R}; z_R,z'_R),
\end{eqnarray}
and
\begin{eqnarray}
\phi_{out}({\bf r_R}) = \frac{1}{2ik_R} \int_{\sqrt (R^2-b^2)}^{\infty} d z'_R
\phi_R
({\bf b_R}; z_R,z'_R).
\end{eqnarray}
Here
\begin{eqnarray}
\phi_R({\bf b_R}; z_R,z'_R) &=& exp[\frac {1}{2ik_R}
(\mu^2 -m^2_R+i\Gamma_R m_R)(z_R-z^{'}_R)] \nonumber \\
& & \times exp[\frac {-i}{v_R}
\int ^{z'_R}_{z_R}
V_R (b_R,z{''}_R) d z^{''}_R].
\end{eqnarray}
After a little bit of manipulations, the final expressions for
$\phi_{in}$ and $\phi_{out}$ work out to ,
\begin{eqnarray}
\phi_{in}({\bf r_R}; k_R,\mu) = \frac {G_0^*} {2 m_R} [1 - exp(\frac {i}
{v_R G_0^*} [L(b_R) - z_R])],
\label{phiin}
\end{eqnarray}
\begin{eqnarray}
\phi_{out}({\bf r_R}; k_R,\mu) = \frac {G_0} {2 m_R} [ exp(\frac {i}
{v_R G_0^*} [L(b_R) - z_R])],
\label{phiout}
\end{eqnarray}
where  $v_R$  is the speed of the resonance and $ L (= \sqrt{(R^2
-b ^2_R)})$ is the  length  from  the  production  point  to  the
surface  of the nucleus. $G_{0}$ and $G_{0}^{*}$ in Eqs. (12) and
(13) are the free and the in-medium resonance propagators.  Their
forms are
\begin{eqnarray}
G_0= \frac{2 m_R} {\mu^2 - m^2_R +i \Gamma _R m_R},
\end{eqnarray}
\begin{eqnarray}
G_0^*= \frac{2 m_R} {\mu^2 - m^{*2}_R +i \Gamma _R^* m_R},
\end{eqnarray}
with
\begin{eqnarray}
m_R^*\approx&  m_R + \frac{E_R}{m_R} U_R .
\end{eqnarray}
\begin{eqnarray}
\Gamma ^* _R = \Gamma_R + \frac{E_R}{m_R} |2 W_R|.
\end{eqnarray}

It  may  be  mentioned  that, for a nucleus with no sharp surface
approximation  the  expression  given  in  Eq.~(6)  can  be  used
directly to evaluate the function $\phi({\bf r_R;k_R,\mu})$.

In the above we have written,
\begin{eqnarray}
V_R= U_R + i W_R.
\end{eqnarray}
These potentials, as given in Eqs.~(16,17) give a measure of the
mass  and  width  modification  of  the  resonance in the nuclear
medium. Their values are an open question and a subject  of  much
research  internationally. In one approach they can be treated as
completely unknown quantities and data on appropriate experiments
can be used to extract their values. This exercise  would  be  of
use  if  the  theoretical  formalism  used describes the reaction
dynamics correctly and the data do  not  have  much  uncertainty.
Alternatively,  they  can  be estimated in a particular model and
the ensuing values can be used to make an estimate of  the  cross
section  for  the  rho production. In literature, various efforts
\cite{rhop,kon} have been made to estimate $V_R$ using  the  high
energy ansatz, i.e.
\begin{eqnarray}
U_R = -\alpha [\frac{1}{2} v_R \sigma _ T^{RN} \rho_0]
\label{uurr}
\end{eqnarray}
and
\begin{eqnarray}
W_R= -[\frac{1}{2} v_R \sigma _ T^{RN} \rho_0],
\label{vvrr}
\end{eqnarray}
where  $\alpha$ is the ratio of the real to the imaginary part of
the elementary RN scattering amplitude and $\sigma _  T^{RN}$  is
the  total  cross section for it. $\rho_0$ is the typical nuclear
density. A detailed calculation for the rho-meson has  been  done
on these  lines  by  Kondrayuk et al. \cite{kon} which give these
potentials as a function  of  momentum.  They  use  VDM  at  high
enegies and resonance model at low energies to generate the $\rho
$N  scattering  parametrs.  We  have  used  these  values for our
calculations in the present paper. Some representative values  of
the self energies required in our calculations are given in Table
1.
\begin{table}
\begin{center}
\caption{Rho-meson optical potentials following Ref.~[7] for
$\mu $=770 MeV.}
\medskip
\begin{tabular} {ccc}
\hline
 $v/c$         & U(MeV) & W(MeV)\\
\hline
0.04    & -20.4     & -40.8  \\
0.6    & 37.9     & -50.6  \\
0.9   & 25.8     & -54.7  \\
\hline
\end{tabular}
\end{center}
\end{table}

\subsection {Results and Discussion}
Examining Eqs. (9-11) $\phi({\bf r_R;k_R,\mu})$ in above we find
that  the  cross  sections  for the decay of the resonance in the
nucleus depends upon the length of the nuclear medium, the  speed
($v_R$),  free  decay  width and self energy of the resonance. To
represent the effect of all these quantities, we present  results
for  the decay of the rho-meson for different values of $v_R$ and
for  two  sets  of  nuclear systems, viz. $Pb+Pb$ and $S+Au$. The
free width of the rho-meson  is  taken  equal  to  150  MeV.  The
optical potentials, which we need at several $\rho $ momenta, are
taken,  as  mentioned  above,  from Kondratyuk et al. \cite{kon}.
Nuclear densities are taken from Ref.~\cite{at}.

Denoting  the  ratio  $\frac  {1}{[K.F.]}  \frac  {\Delta  \sigma
_R^{AB}}{\Delta \sigma _R^{NN}}$ in Eq. (2) as $|\Phi|^2$, we  plot
$|\Phi|^2$  as  a  function  of the invariant mass, $\mu$, of the
decay  products  of  the  $\rho$  meson.  Figures  1-4  show  the
invariant  mass spectra of the $\rho$ meson in $Pb+Pb$ and $S+Au$
collisions  at  rho  velocities of $0.6c$ and $0.9c$. 
\begin{figure}
\hbox{\hspace{6em}
\hbox{\psfig{figure=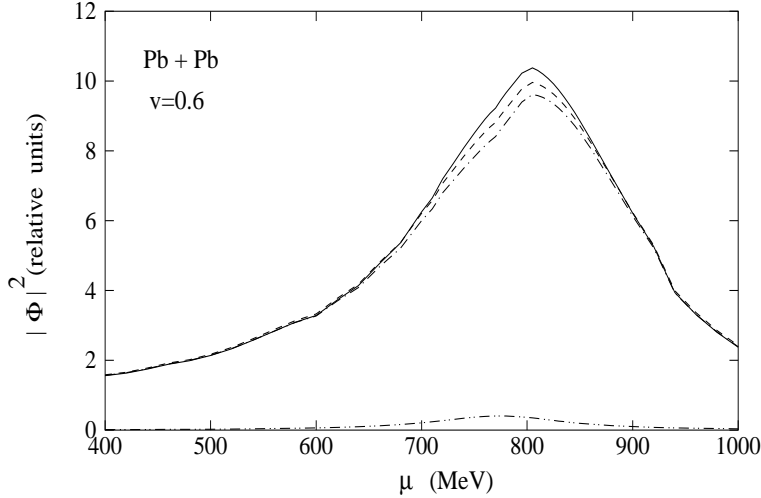,height=6.5 cm,width=10.0 cm}}}
\caption{ The invariant mass spectra of the $\rho$ meson produced
in Pb+Pb collisions. The solid curve is obtained after adding the
inside and the ouside decay coherently, while the dashed curve is
obtained  by  adding the same incoherently. The dash-dot curve is
the inside decay contribution and the dash-dot-dot curve  is  the
outside  decay  contribution  separately.  The resonance speed is
0.6c.}
\label{fig1}
\end{figure}
\begin{figure}
\hbox{\hspace{6em}
\hbox{\psfig{figure=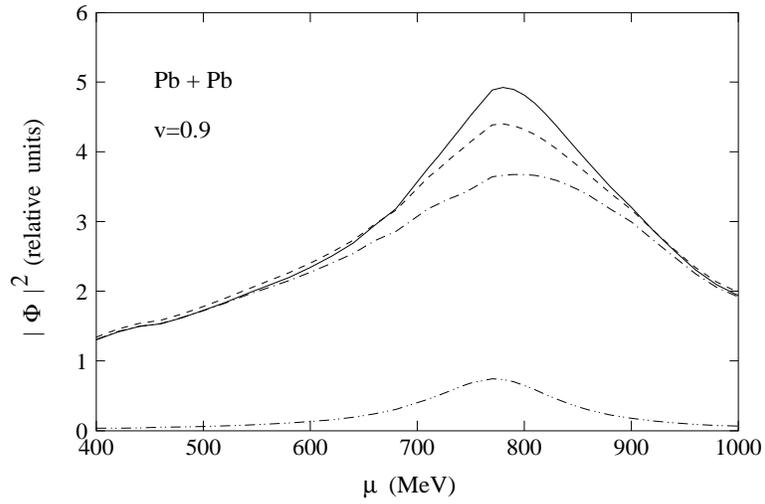,height=6.5 cm,width=10.0 cm}}}
\caption{  All  curves  have  same  meaning  as  in  Fig.~1.  The
resonance speed is 0.9c.}
\label{fig2}
\end{figure}
\begin{figure}
\hbox{\hspace{6em}
\hbox{\psfig{figure=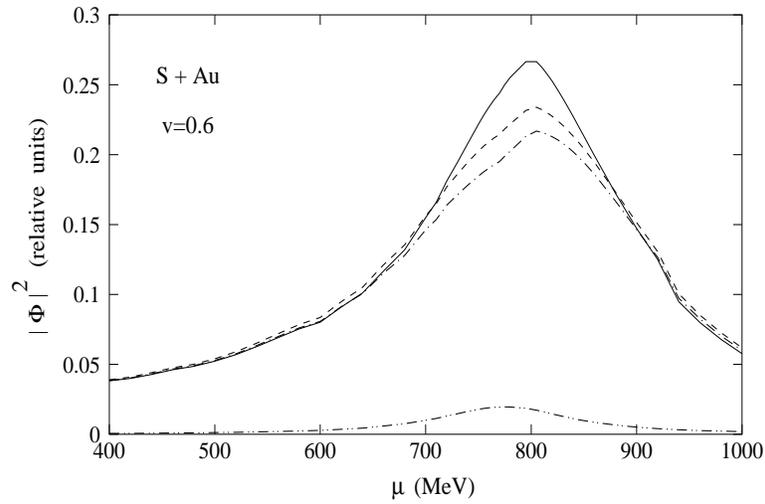,height=6.5 cm,width=10.0 cm}}}
\caption{  All  curves  have  same  meaning  as  in  Fig.~2.  The
colliding nuclear system is  S+Au  and  the  resonance  speed  is
0.6c.}
\label{fig3}
\end{figure}
\begin{figure}
\hbox{\hspace{6em}
\hbox{\psfig{figure=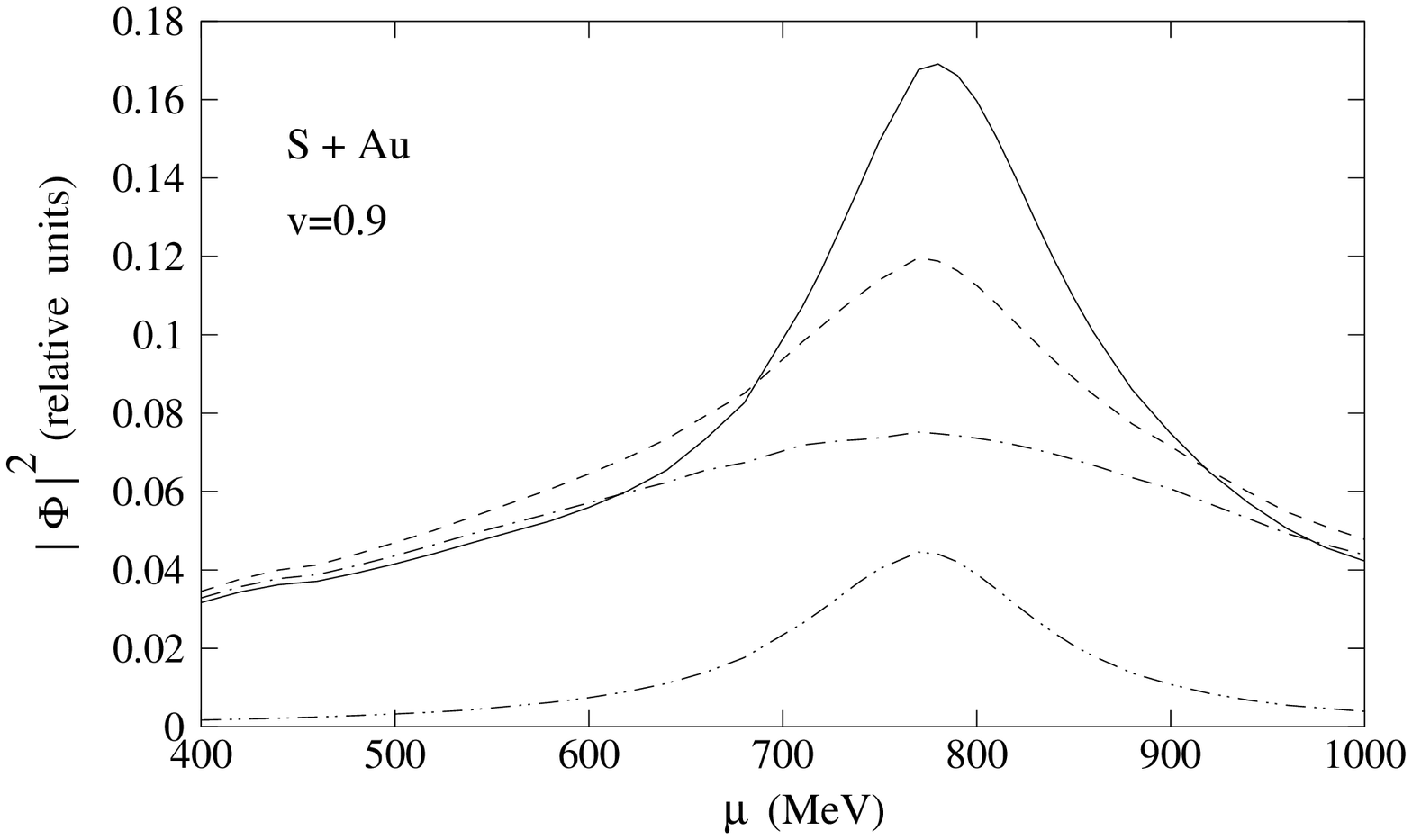,height=6.5 cm,width=10.0 cm}}}
\caption{ Same as in Fig.~3. The resonance speed is 0.9c.}
\label{fig4}
\end{figure}
Here $c$ is
the speed of light. The solid curve in all the  figures gives the
coherently  summed  cross-section from the decay of $\rho $-meson
inside and outside the nuclear medium. The dashed curve gives the
same   added   incoherently.   The    individual    contributions
corresponding  to  the  inside and the outside decay are given by
the dash-dot and dash-dot-dot curves respectively. We observe two
things. One, the coherent and the incoherent  cross-sections  are
different and second, this difference increases with the increase
in  the  rho-meson  speed.  At 0.6c speed, while the coherent and
incoherent curves differ only in the peak cross sections, at 0.9c
speed their shape and peak cross sections both are different.  We
also observe that the difference is larger for the smaller system
like S on Au.
If we compare the mass shift seen in our calculations (Figs.~1-4)
with those indicated in the high energy heavy ion collisions, our
shifts are small and are in the opposite direction. To explore as
what kind of optical potentials would produce a shift as large as
those indicated experimentally we calculated $|\Phi|^2$ for Pb+Pb
at  0.9c  for three arbitrarily chosen values of $U_R$, viz. -40,
-80 and -120 MeV. These results are shown in Fig.~5. 
\begin{figure}
\hbox{\hspace{6em}
\hbox{\psfig{figure=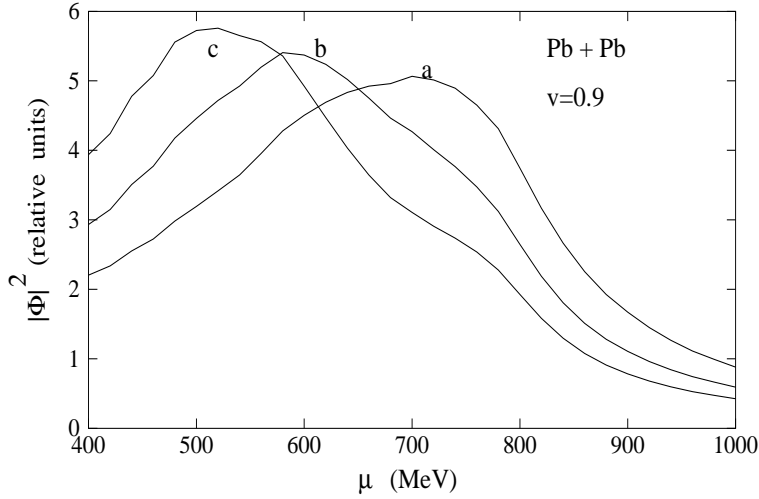,height=6.5 cm,width=10.0 cm}}}
\caption{  Sensitivity of the cross-section to the amount of mass
modification of the rho-meson. The calculations are presented for
Pb+Pb system at 0.9c rho-meson speed. Curves 'a', 'b' and 'c' are
for $U_R$ equal to -40, -80 and -120 respectively. The  imaginary
part of the optical potentials is the same as used in Fig.~2. }
\label{fig5}
\end{figure}
We find that only with -120 MeV value the distribution starts 
having  features
resembling  those  indicated  in  heavy-ion experiments. But this
value, compared with the value around +25  MeV  coming  from  the
high  energy  ansatz  of  Kondratyuk et al. (see Table 1) is very
large and is of opposite sign.

\section{Proton-nucleus collisions}
The  cross section for a coherent rho production reaction,
(p,p$^\prime \rho^0$), is given by
\begin{equation}
d \sigma = [PS]  S(m^2) <|T_{coh}|^2>,
\end{equation}
where the phase-space factor, [$PS$], is written as
\begin{equation}
[PS]=\frac{\pi m^2_p m_A}{(2\pi)^6} \frac{k_{p^\prime}k^2_{\rho}}
 {k_p[k_\rho(E_i-E_{p^\prime})-{\bf (k_p-k_{p^\prime})}.{\hat{k_\rho}}E_\rho]}
 dm^2dE_{p^\prime}d\Omega_{p^\prime}d\Omega_\rho.
\end{equation}
$S(m^2)$  is the free space rho mass distribution function, which
is given by
\begin{equation}
S(m^2)=\frac{1}{\pi} \frac{m_\rho \Gamma_\rho}
       {[(m^2-m_\rho^2)^2 + m_\rho^2 \Gamma_\rho^2]},
\end{equation}
with $m_\rho =770$ MeV and $\Gamma_\rho =150$ MeV.

The T-matrix, $T_{coh}$, is given by
\begin{equation}
T_{coh}=(\chi^{(-)*}_{\bf k_{p^\prime}}, \Psi^{(-)*}_{\bf k_\rho}
<p^\prime,\rho^0|{\cal L}_{\rho NN}|p>\chi^{(+)}_{\bf k_p}),
\end{equation}
where  $\chi  $'s denote the distorted waves for the incoming and
outgoing protons. However, in the energy region of  interest  for
rho  production  the  distortion  effects  are mainly absorptive.
Therefore, the proton distorted waves in above can be replaced by
plane  waves for the present purpose.

$\Psi^{(-)*}_{\bf  k_\rho}$  is the $\rho $-meson scattering wave
function with asymptotic momentum ${\bf k_\rho  }$.  It  has  the
form
\begin{equation}
\Psi^{(-)*}_{\bf k_\rho }=e^{-i{\bf k_\rho \cdot r}}+ \Psi_{scat.}^*
\end{equation}
In  the  absence  of  any  dispersive nuclear distortion of p and
p$^\prime$, the first term in this equation does  not  contribute
to   $T_{coh}$  because  ${\bf  k_\rho}  \not  =  {\bf  k_p}-{\bf
k_{p^\prime}}$. This in other words  means  that  the  rho  meson
produced at the proton vertex is off-shell. It can not be seen in
the  detector  without  incorporating the medium effects on it. $
\Psi_{scat. }$ is the part of the  wave  function  which  include
these  effects.  If we associate a self energy $\Pi (=2\omega V$,
where $V$ is the corresponding optical potential) with the  $\rho
$-meson, $\Psi_{scat.}$ is given by
\begin{equation}
\Psi^*_{scat.} = \chi^{(-)*}_{\bf k_\rho} V G_\rho (t),
\end{equation}
where  $\chi^{(-)*}_{\bf  k_\rho}$  is the scattering solution of
the potential $V$. $G_\rho(t)$ is  the  $\rho$-meson  propagator,
and is given by
\begin{equation}
G_\rho (t) = -\frac{2\omega}{m^2_\rho -t-i\omega\Gamma_\rho}.
\end{equation}
$t(=\omega^2-{\bf q}^2)$ is the four-momentum  transfer squared
to $\rho$-meson at the production vertex.

For the $\rho $ production Lagrangian in Eq.~(24) we have taken
\begin{equation}
{\cal L}_{\rho NN}=\frac{fF(t)}{m_\rho}
                    N^{\dag} {\bf (\sigma x q) \tau}N \cdot
                   {\bf \rho},
\end{equation}
with  $\rho$NN  coupling  constant,  $f$,  equal to 7.81, and the
off-shell extrapolation form factor as
\begin{equation}
F(t)=\frac{\Lambda^2-m_\rho^2}{\Lambda^2-t},
\end{equation}
with $\Lambda$= 2 GeV/c.

With  the  above  formalism  we  calculate the $\rho $ production
cross section for the $^{12}$C target nucleus. The only  quantity
required  for  the  calculation is the description of the optical
potential, $V$, of the $\rho$-meson. The values  of  the  optical
potential  are fixed using the same prescription as given earlier
for the heavy ion reactions. Some representative values  required
by  us  are  given  in  Table~2.  The radial shape of the optical
potential is approximated by the radial density  distribution  of
the $^{12}$C nucleus.

\begin{table}
\begin{center}
\caption {Optical potentials for certain values of the $\rho$
momentum.}
\medskip
\begin{tabular}{ c  c  c  c  c  c }
\hline
k$_\rho$ (MeV/c) &   30   &   50   &  75    &   100     &   500    \\
\hline
   U (MeV)       & -20.75 & -9.27  & -0.37  &   5.90    &   33.49  \\
   W (MeV)       & -26.51 & -30.46 & -34.79 &  -42.09   &  -46.73  \\
\hline
\end{tabular}
\end{center}
\end{table}

\begin{figure}[h]
\hbox{\hspace{6em}
\hbox{\psfig{figure=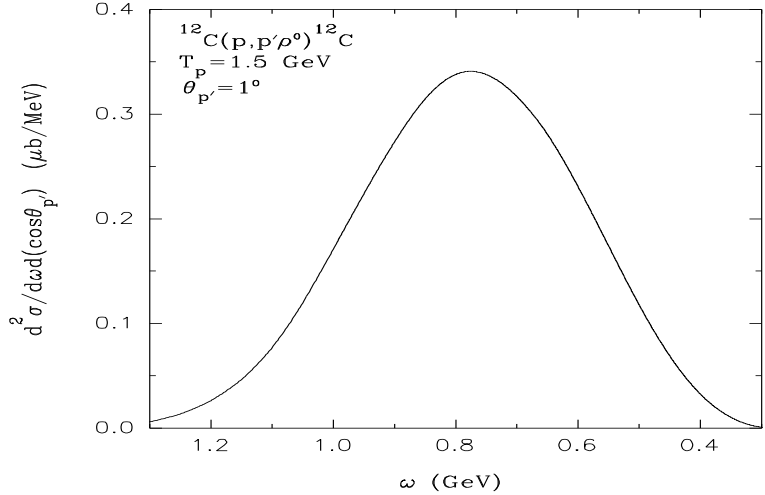,height=6.5cm,width=10.0 cm}}}
\caption{ Calculated energy spectrum for the outgoing proton near
forward  angle  as  a  function  of  the  energy transfer $\omega
(=\mbox{T}_p  -  \mbox{T}_{p^\prime})$  for  $^{12}$C(p,p$^\prime
\rho^0)^{12}$C  reaction  integrated  over all emission angles of
the $\rho^0$-meson. The beam energy is 1.5 GeV  and  the  optical
potentials are given in Eqs.~(\ref{uurr}) and (\ref{vvrr}).}
\label{kk1}
\end{figure}

In Fig.~\ref{kk1} we  plot the calculated outgoing proton energy
spectrum for p$^\prime $ going very near to the forward direction
against the  energy  transfer  $\omega  $(=T$_p$-T$_{p^\prime}$).
This energy transfer and the corresponding momentum transfer {\bf
q}(=${\bf  k_p}$  -  ${\bf k_{p^\prime}}$) are shared between the
rho-meson and the recoiling nucleus through  the  interaction  of
the  rho-meson  with the target nucleus. The beam energy is taken
equal to 1.5 GeV. We  see  in  the  figure  that  the  calculated
distribution  has  a broad peak. The peak cross section is around
0.34 $\mu$b/MeV.
\begin{figure}
\hbox{\hspace{6em}
\hbox{\psfig{figure=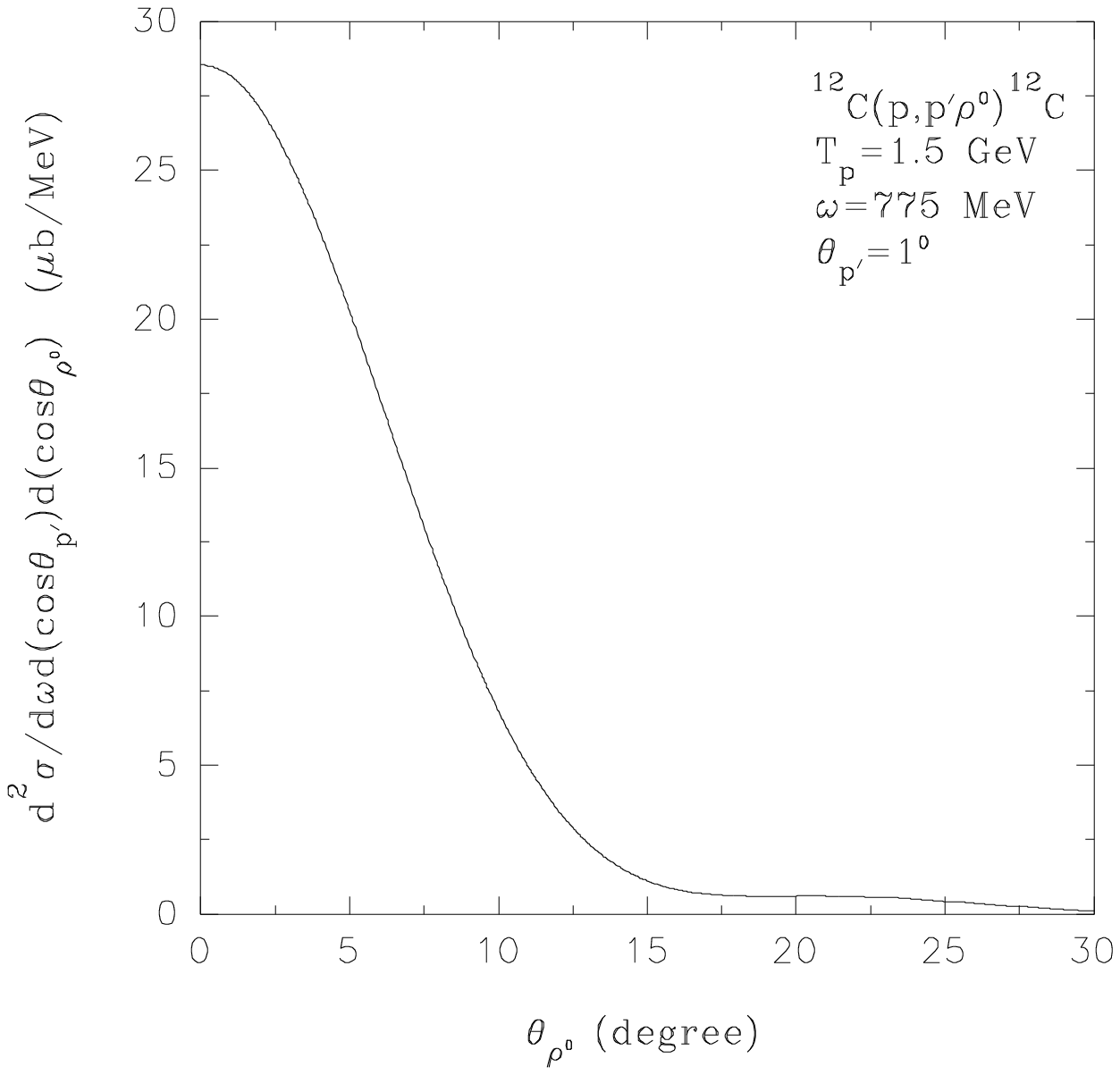,height=6.5cm,width=10.0 cm}}}
\caption{Calculated  angular distribution of the $\rho^0$-meson
for the energy transfer ($\omega$) at the peak in Fig.~\ref{kk1}.}
\label{kk2}
\end{figure}

In  Fig.~\ref{kk2}  we show the angular distribution of the above
rho-mesons at the peak position in Fig~\ref{kk1}. It is  observed
that  most  of  the  rho-meson  flux  gets emitted in the forward
direction only. Very little is seen beyond 15$^0$ or so.
\begin{figure}
\hbox{\hspace{6em}
\hbox{\psfig{figure=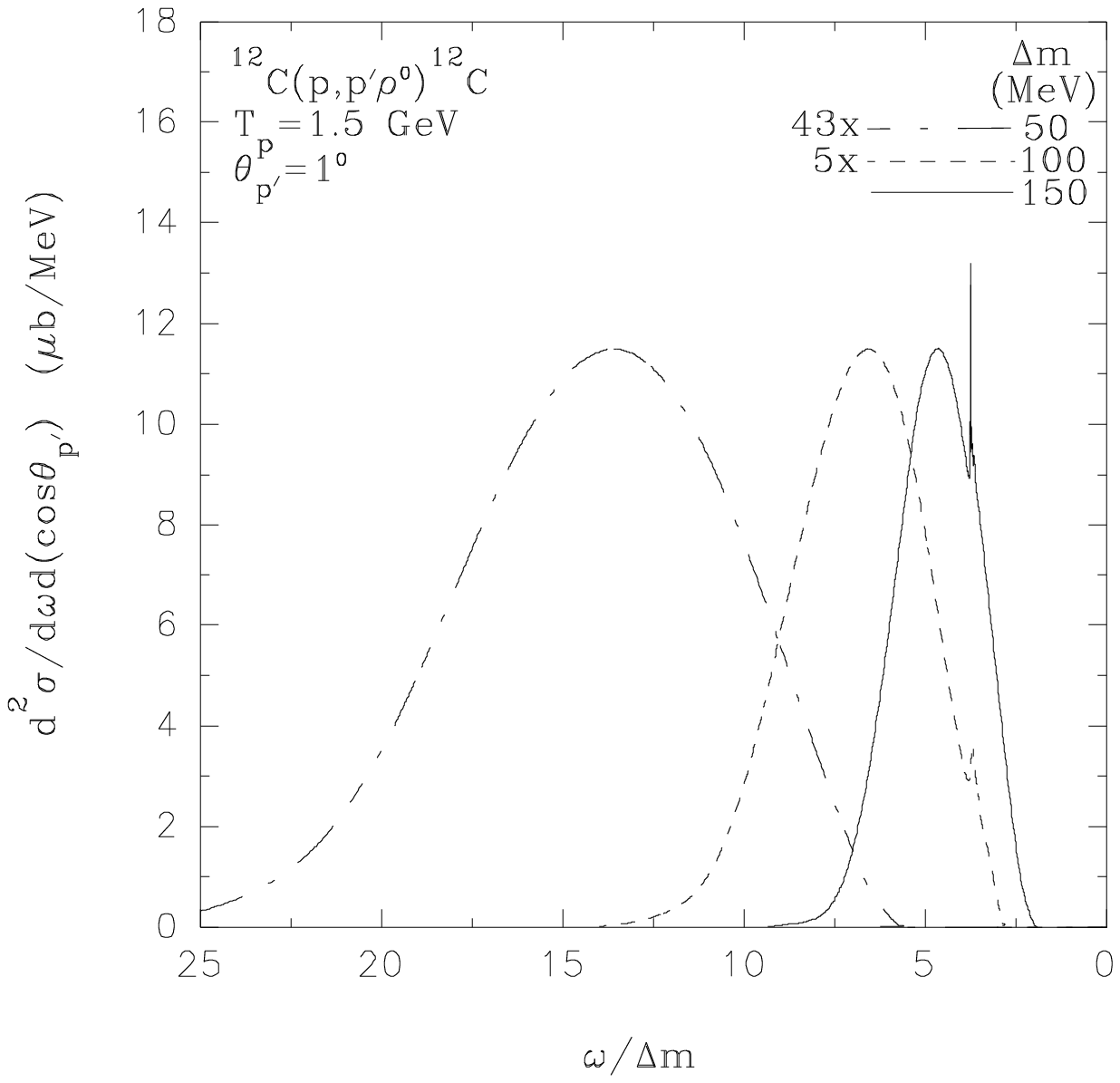,height=6.5cm,width=10.0 cm}}}
\caption{ Calculated energy spectrum for the outgoing proton near
forward  direction as a function of the energy transfer $\omega $
in the units of  $\rho^0$-meson mass-shift, $\Delta m$, at 1.5
GeV beam energy for different values of the mass-shift. }
\label{jj1}
\end{figure}

Above  results are given for a certain choice of the $\rho$-meson
optical potential. However, they would be sensitive to the change
in this potential. In Fig.~\ref{jj1} we  have  investigated  this
sensitivity.  The  optical  potential  for  this purpose has been
taken purely real, and different values for it are fixed  through
different mass-shifts of the $\rho$-meson in the medium using the
relation  given  in  Eq.~(16).  In  Fig.~\ref{jj1}  we  show  the
calculated proton energy spectrum for  $\Delta  \mbox{m(=m-m}^*)$
taken equal to 50, 100 and 150 MeV. On x-axis, instead of $\omega
$,  we  have $\frac{\omega }{\Delta m}$. This is done because the
essential parameter determining the dynamics of the rho-meson  in
the  potential  is  likely  to  be the rho energy relative to the
depth of the potential. We observe that

\begin{enumerate}
\item  the  magnitude  of  the  cross sections increases with the
increase in the strength of the potential.

\item  In  addition to the broad peak, we see a sharp peak in the
small energy region of the rho-meson. The position of  this  peak
on  the  $\frac{\omega}  {\Delta  m}$  scale  is  around  3.7 for
$\Delta$m=100 and 150 MeV. For $\Delta $m=50 MeV,  this  peak  is
not seen in the results because by then the cross section becomes
too small.

On  examining  the phase-shifts of the scattered wave function of
rho-meson in the potential, we find that the sharp peak is like a
shape  elastic  resonance  seen   in   the   elastic   scattering
experiments.

Of  course,  when  the $\rho$-potential is made complex, as is in
Fig.~\ref{kk1} the sharp peak disappears.
\end{enumerate}

To  summarize,  we  find  that  in  proton scattering on nuclei a
measurable cross section exists for $\rho $ meson production  due
to coherent effect of the target nucleus. The actual magnitude of
the  cross  section  depends  sensitively  on the strength of the
$\rho $-meson optical potential, which is related to the rho-mass
modification in the nuclear medium. The cross  section  increases
with   the  increase  in  the  potential  strength.  The  angular
distribution of the emitted rho-meson is such that most  of  them
go in a forward cone of about 15$^0$. For a purely real potential
a  sharp peak appears in the proton energy spectrum in the region
of the small rho-meson energy.\\\\

The  authors  acknowledge  many  useful discussions they had with
Shashi Phatak and A. B. Santra, and thank them for the same.

\end{document}